\documentclass[twocolumn, twocolappendix, numberedappendix, appendixfloats]{openjournal}
\usepackage{xcolor}
\usepackage{textgreek}
\usepackage[utf8]{inputenc}
\usepackage[english]{babel}

\usepackage{hyperref}
\hypersetup{
    unicode,
    colorlinks=true,
    linkcolor=linkcolor,
    citecolor=linkcolor,
    filecolor=linkcolor,
    urlcolor=linkcolor,
}
\usepackage{color,colortbl}
\definecolor{linkcolor}{rgb}{0.0,0.3,0.5}
\usepackage{tensind}
\tensordelimiter{?}
\DeclareGraphicsExtensions{.bmp,.png,.jpg,.pdf}
\usepackage{verbatim}
\usepackage[normalem]{ulem}
\usepackage{orcidlink}
\usepackage{soul}

\DeclareRobustCommand{\VAN}[3]{#2}
\let\VANthebibliography\thebibliography
\def\thebibliography{\DeclareRobustCommand{\VAN}[3]{##3}\VANthebibliography}

\usepackage{graphicx}	% Including figure files
\usepackage{adjustbox}  % For the orcid ids
\usepackage{amsmath}	% Advanced maths commands
\usepackage{wasysym}    % Extra astronomy symbols
\usepackage{bm}         % Bold maths
\usepackage{xspace}
\usepackage{multirow}
\makeatletter % Workaround to only color the year for cite commands
  \patchcmd{\NAT@citex}
    {\@citea\NAT@hyper@{%
      \NAT@nmfmt{\NAT@nm}%
      \hyper@natlinkbreak{\NAT@aysep\NAT@spacechar}{\@citeb\@extra@b@citeb}%
      \NAT@date}}
    {\@citea\NAT@nmfmt{\NAT@nm}%
    \NAT@aysep\NAT@spacechar\NAT@hyper@{\NAT@date}}{}{}

  \patchcmd{\NAT@citex}
    {\@citea\NAT@hyper@{%
      \NAT@nmfmt{\NAT@nm}%
      \hyper@natlinkbreak{\NAT@spacechar\NAT@@open\if*#1*\else#1\NAT@spacechar\fi}%
        {\@citeb\@extra@b@citeb}%
      \NAT@date}}
    {\@citea\NAT@nmfmt{\NAT@nm}%
    \NAT@spacechar\NAT@@open\if*#1*\else#1\NAT@spacechar\fi\NAT@hyper@{\NAT@date}}
    {}{}
\makeatother \usepackage{amssymb}	% Extra maths symbols

\newcommand\Msun{\text{M}_{\astrosun}}

\DeclareRobustCommand{\HI}{{\text{H}\,\textsc{i}}\xspace} % neutral hydrogen
\DeclareRobustCommand{\HII}{{\text{H}\,\textsc{ii}}\xspace} % ionized hydrogen
\DeclareRobustCommand{\HeI}{{\text{He}\,\textsc{i}}\xspace} % neutral helium
\DeclareRobustCommand{\HeII}{{\text{He}\,\textsc{ii}}\xspace} % singly ionized helium
\DeclareRobustCommand{\thesan}{\mbox{\textsc{thesan}}\xspace}

\DeclareRobustCommand{\lumina}{\mbox{\textsc{lumina}}\xspace}
\DeclareRobustCommand{\thesanone}{\mbox{\textsc{thesan-1}}\xspace}
\DeclareRobustCommand{\thesantwo}{\mbox{\textsc{thesan-2}}\xspace}

\DeclareRobustCommand{\arepo}{\mbox{\textsc{arepo}}\xspace}
\DeclareRobustCommand{\areport}{\mbox{\textsc{arepo-rt}}\xspace}

\newcommand\orcid[1]{\href{http://orcid.org/#1}{\adjustbox{trim={-.15\width} {0\height} {-.15\width} {0\height},clip}{\includegraphics[height=10pt]{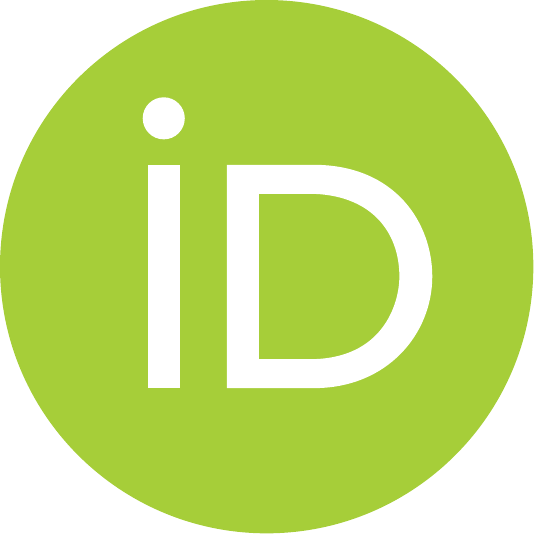}}}}

\newlength{\myfigwidth}
\graphicspath{ {./ws_figures/} }

\begin{document}
\setlength{\myfigwidth}{\linewidth}
\title{The Lumina Project: Morphology of Ionized Bubbles and Neutral Islands}

\author{Meredith Neyer\orcid{0000-0002-9205-9717}$^{1\,\star}$}
\email[$^\star$ E-mail: \href{mailto:mneyer@mit.edu}{mneyer@mit.edu}]{}
\author{Aaron~Smith\orcid{0000-0002-2838-9033}$^{2}$}
\author{Ali~Sadain\orcid{0009-0009-5786-7469}$^{2}$}
\author{Oliver~Zier\orcid{0000-0003-1811-8915}$^{3}$}
\author{Mark~Vogelsberger\orcid{0000-0001-8593-7692}$^{1,4}$}
\author{Teodora-Elena~Bulichi\orcid{0000-0001-8174-6389}$^{1}$}
\author{Xuejian~Shen\orcid{0000-0002-6196-823X}$^{1}$}
\author{Sonja~M.~Koehler\orcid{0009-0008-0814-3328}$^{3}$}
\author{Volker~Springel\orcid{0000-0001-5976-4599}$^{5}$}
\author{Rahul~Kannan\orcid{0000-0001-6092-2187}$^{6}$}
\author{Rongrong~Liu\orcid{0000-0003-0685-3525}$^{3}$}
\author{Lars~Hernquist\orcid{0000-0001-6950-1629}$^{3}$}
\author{Changhyun~Cho\orcid{0000-0002-9879-1749}$^{6}$}
\affiliation{$^{1}$Department of Physics $\&$ Kavli Institute for Astrophysics and Space Research, Massachusetts Institute of Technology, Cambridge, MA 02139, USA}
\affiliation{$^{2}$Department of Physics, The University of Texas at Dallas, Richardson, Texas 75080, USA}
\affiliation{$^{3}$Center for Astrophysics $|$ Harvard $\&$ Smithsonian, 60 Garden Street, Cambridge, MA 02138, USA}
\affiliation{$^{4}$Fachbereich Physik, Philipps Universit\"at Marburg, D-35032 Marburg, Germany}
\affiliation{$^{5}$Max Planck Institute for Astrophysics, Karl-Schwarzschild-Str.\ 1, D-85741 Garching, Germany}
\affiliation{$^{6}$Department of Physics and Astronomy, York University, 4700 Keele Street, Toronto, ON M3J 1P3, Canada}

\begin{abstract}
The Epoch of Reionization (EoR) is characterized by the nucleation, expansion, and percolation of ionized bubbles in the intergalactic medium (IGM). Interpreting emerging multi-wavelength observations from facilities such as \textit{JWST} and next-generation 21\,cm interferometers requires a detailed theoretical understanding of the 3D morphology of these \HII regions and of the residual neutral structures that survive near the end of reionization. We analyze the geometry and physical properties of ionized bubbles and late-time neutral islands in the large-volume $(500\,\text{cMpc})^3$ radiation-hydrodynamics simulation \lumina. We identify discrete structures with a 3D watershed method and measure their volumes, shape tensors, and internal gas properties. The volume-weighted bubble size distributions evolve substantially across the EoR: characteristic radii shift from $R_\text{eff} \sim 1$\,cMpc at $z \gtrsim 10$ to $R_\text{eff} \gtrsim 100$\,cMpc by $z \lesssim 6$, while watershed-derived sizes exceed line-of-sight mean-free-path (MFP) measurements by roughly a factor of three at intermediate redshifts. The bulk bubble population is predominantly prolate through most of the EoR. By $z \approx 6$, however, the volume-weighted distribution of the triaxiality parameter $T$, which distinguishes flattened from elongated structures, becomes distinctly bimodal in the fiducial segmentation, with a sharp oblate peak near $T \approx 0.25$ associated with a small number of very large bubbles during late percolation. The properties of bubble interiors are consistent with inside-out reionization, with median gas overdensities ($\delta$) within bubbles decreasing from $\delta \sim 1$ at $z = 15$ toward the cosmic mean by $z = 6$. Beyond a minimum near the resolution scale, the median internal UV radiation energy density increases with bubble size. The residual neutral islands at $z < 6$ peak in volume-weighted size near $R_\text{eff} \approx 10$\,cMpc and remain predominantly prolate, consistent with anisotropic residual neutral structure near the end of reionization.
\end{abstract}


\maketitle

\section{Introduction}
\label{sec:intro}

The Epoch of Reionization (EoR) marks a major transition in the Universe, during which ultraviolet radiation from the first generations of stars and galaxies ionized the neutral hydrogen in the intergalactic medium \citep[IGM;][]{Shapiro1987, BarkanaLoeb2001, Furlanetto2006b, Wise2019}. This process begins with the formation of discrete ionized regions, or ``bubbles'', around early luminous sources. As the EoR progresses, these localized \HII regions expand and eventually merge through a complex percolation process. Observational constraints suggest that reionization is largely complete by $z \approx 5.3$--$5.7$, while models still allow residual neutral structure to persist to somewhat lower redshift \citep{Fan2006, Bosman2022}. Understanding the timeline and morphological evolution of the bubbles is therefore important both for constraining the early ionizing sources and for interpreting observations from current and next-generation facilities, including the \textit{James Webb Space Telescope} (\textit{JWST}) and 21\,cm radio interferometers such as the Low Frequency Array \citep[LOFAR;][]{vanHaarlem2013}, Hydrogen Epoch of Reionization Array \citep[HERA;][]{DeBoer2017, HERA2023}, and Square Kilometre Array \citep[SKA;][]{Mellema2013}.

Directly constraining the physical scale and morphology of these \HII regions remains observationally challenging, and recent multi-wavelength campaigns approach the problem through several complementary tracers. Upper limits on the 21\,cm power spectrum constrain models of reionization and IGM heating because spatial variations in the 21\,cm brightness temperature depend on both the ionization topology and the thermal state of the gas \citep[e.g.,][]{Abdurashidova2022}. At the same time, spectroscopy of $z \gtrsim 6$ quasars probes proximity zones whose transmission depends on the quasar luminosity and lifetime as well as the surrounding IGM \citep[e.g.,][]{Eilers2017}. Beyond these zones, Ly$\alpha$ forest opacity constrains the late stages of reionization \citep{Bosman2022}. The neutral IGM also imprints a characteristic Ly$\alpha$ damping wing on the spectra of high-redshift sources, with the strength of the absorption depending on the distance to the neutral-gas boundary \citep{Keating2023a, Mason2026}. These observables probe related aspects of the reionization topology, but they do not weight the three-dimensional extent, connectivity, and distance to a neutral boundary in the same way.

Measurements of Ly$\alpha$ transmission from Lyman-$\alpha$ emitters (LAEs) and damping-wing absorption in galaxy spectra can therefore constrain the physical extent of local ionized bubbles \citep[e.g.,][]{Mason2018, Umeda2023, Neyer2026}. Ground-based spectroscopy and targeted \textit{JWST} programs have characterized the local environments of LAEs at roughly $5.8 \lesssim z \lesssim 8.7$, finding enhanced Ly$\alpha$ transmission in several galaxy overdensities, consistent with large ionized regions around clustered sources \citep[e.g.,][]{Endsley2022, Larson2022, Witstok2024}. Recent Ly$\alpha$-based mapping methods have also been developed to infer bubble edges, locations, and characteristic sizes from ensembles of galaxies, making estimator-specific calibration increasingly important \citep{Lu2025, Nikolic2025}. Although these empirical techniques constrain the timeline and characteristic scales of the EoR, interpreting the different tracers together requires theoretical models that connect their observable signatures to the underlying 3D topology of the IGM.

Large-scale radiation-hydrodynamics (RHD) simulations provide a means of following the coupled, highly non-linear processes that drive reionization within a single numerical framework. The \lumina project belongs to this class of calculations, combining the IllustrisTNG galaxy-formation model with the \areport radiative-transfer solver to produce a self-consistent, high-resolution model of the EoR \citep{Zier2026}. RHD studies have characterized the global reionization history and the statistical properties of the IGM in considerable detail, but the complex and evolving geometry of individual ionized bubbles remains difficult to quantify.

Bubble sizes have been measured using a variety of methods, including the mean-free path \citep[MFP;][]{Mesinger2007}, spherical averaging \citep[SPA;][]{Zahn2007}, friends of friends \citep[FOF;][]{Iliev2006,Friedrich2011,Iliev2014}, granulometry \citep{Kakiichi2017}, and watershed \citep{Lin2016} methods. A key lesson from estimator comparisons is that ``bubble size'' is not a unique, method-independent observable, because different algorithms weight connectedness, cross-section, and line-of-sight path length differently \citep{Lin2016, Lu2024}. Bubble sizes quantified using the MFP method define characteristic scales via averages of ray lengths extending from ionized cells to the first neutral boundary along each ray, where the boundary is defined operationally by the first crossing into cells below a chosen ionization threshold rather than by a unique physical surface \citep[e.g.,][]{Mesinger2007, Lin2016, Neyer2024}. This method characterizes ionized path lengths relevant to the transmission of light from high-$z$ galaxies. Watershed segmentation instead identifies connected basins and measures the volumes of individual \HII structures while retaining information about their topology \citep{Lin2016, Lu2024}. As reionization progresses, bubbles do not expand isotropically, and their growth is modulated by the underlying cosmic web, leading to highly complex, non-spherical geometries \citep{Furlanetto2004a, McQuinn2007a}.

To capture this topological complexity, we use a 3D watershed segmentation method to identify and isolate individual ionized bubbles in the large-volume \lumina simulation. In contrast to a single line-of-sight length, the watershed boundaries retain the full connected three-dimensional structure of each region. They therefore allow us to measure bubble volumes and morphologies directly and, in turn, to examine both the growth of the bubbles and the evolution of their shapes from predominantly prolate (filamentary) structures to more complex configurations during the final stages of percolation.

This paper is organized as follows. In Section~\ref{sec:methods}, we detail the numerical methodology, including an overview of the \lumina simulation and our 3D watershed segmentation algorithm. In Section~\ref{sec:sizes}, we present the resulting bubble size distributions across the EoR. Section~\ref{sec:shapes} is dedicated to the morphological analysis of the segmented bubbles, where we quantify the evolution of their triaxiality as a function of redshift and physical scale. Section~\ref{sec:interiors} analyzes the gas properties of bubble interiors, while Section~\ref{sec:neutral} applies the same framework to the residual neutral islands at the close of reionization. We discuss the estimator and volume implications in Section~\ref{sec:discussion} and summarize our primary conclusions in Section~\ref{sec:conclusions}.

\section{Methods}
\label{sec:methods}

In this work, we study the three-dimensional topology of hydrogen reionization in \lumina using high-cadence Cartesian renderings of the ionization field together with a watershed-based bubble finder. Throughout this paper, ``ionized bubbles'' refer to connected \HII regions identified from the local volume-weighted hydrogen ionized fraction, $x_\text{\HII}$.

\subsection{Simulation overview}

The \lumina simulation is a large-volume cosmological RHD calculation designed to follow the coupled evolution of galaxies, black holes, and the IGM through hydrogen and helium reionization down to $z=3$ \citep{Zier2026}. The flagship run evolves a comoving cube of side length $L_\text{box} = 500\,\text{cMpc}$ with $2\times 6000^3$ resolution elements (gas cells and dark matter particles), corresponding to baryonic and dark-matter mass resolutions of $m_\text{gas} = 3.6\times10^6\,\Msun$ and $m_\text{DM} = 1.9\times10^7\,\Msun$. Gravitational forces employ Plummer-equivalent softening lengths of $\epsilon_\text{gas,min} = 0.44\,\text{ckpc}$ for gas (minimum adaptive value) and $\epsilon_\text{DM} = 1.77\,\text{ckpc}$ for dark matter. The adopted cosmological parameters are consistent with the Planck 2018 baseline TT,TE,EE+lowE+lensing+BAO constraints under a $\Lambda$CDM model ($\Omega_m = 0.3096$, $\Omega_b = 0.04897$, $H_0 = 67.66\,\text{km}\,\text{s}^{-1}\,\text{Mpc}^{-1}$, $\sigma_8 = 0.8102$, $n_s = 0.9665$), and the initial redshift is $z_\text{init} = 49$ \citep{Planck2020}.

\lumina is run with the moving-mesh code \arepo \citep{springel2010pur,pakmor2016improving,weinberger2020arepo} and the radiative-transfer module \areport \citep{Kannan2019}, which solves the moment equations of radiative transfer with an M1 closure \citep{Levermore1984,Dubroca1999}. The overall code base incorporates scalability improvements developed for very large cosmological volumes together with GPU acceleration \citep{Gadget4,pakmor2023millenniumtng,zier2024adapting}. For comparison, the complementary \thesanone simulation provides fully coupled RHD in a smaller $95.5$\,cMpc volume with $5.8\times10^5\,\Msun$ gas resolution \citep{Kannan2022a,Garaldi2022,Garaldi2024,Smith2022}. The larger \lumina volume improves the sampling of rare objects and large-scale environments and allows their cosmic variance to be quantified \citep{Zier2026}. This advantage is most relevant to the late-time tail of the bubble distribution, when a small number of large, highly connected structures dominate the ionized volume. Sub-volume tests in \thesan show little variation in the global reionization history \citep{Garaldi2022}, while \lumina reveals significant variation in reionization timing across 125 different $(100\,\text{cMpc})^3$ sub-volumes. Here, the larger volume extends the earlier \thesan bubble results into the regime of the largest percolated structures and rare environments. This regime matters for environmental reionization histories, where overdense regions ionize earlier and with substantial local scatter \citep{Zhao2026}, and for line-intensity mapping, where the topology of ionized and neutral structure controls large-scale Ly$\alpha$, nebular-line, and 21\,cm fluctuations and their cross-correlations \citep{Kannan2022b,Almualla2026}.

\subsection{Radiation hydrodynamics and galaxy formation}

The hydrodynamic equations are evolved on an unstructured, moving Voronoi mesh using a second-order finite-volume Godunov scheme \citep{springel2010pur,pakmor2016improving}. The quasi-Lagrangian mesh motion reduces advection errors, preserves Galilean invariance, and naturally adapts the spatial resolution to the local flow \citep{vogelsberger2012moving}. Time integration follows a hierarchical time-step scheme with approximately constant target gas-cell masses, while self-gravity between gas, dark matter, stars, and black holes is computed with an adaptive Tree--PM solver \citep{Bagla2002,Bagla2003,Barnes1986,Aarseth2003}.

Galaxy formation follows the IllustrisTNG model \citep{Weinberger2017,Pillepich2018Model}, which is an update of the Illustris model \citep{Vogelsberger2013,IllustrisIntro,IllustrisNature}. In brief, star formation proceeds stochastically in cold, dense gas treated with an effective two-phase interstellar medium (ISM) equation of state \citep{Springel2003}, stellar feedback is implemented with a kinetic-wind model, and black holes are seeded in sufficiently massive halos and evolve through accretion, mergers, and feedback \citep{Weinberger2017,Pillepich2018Model,Bulichi2025}. \lumina adopts this galaxy-formation framework, couples it to on-the-fly radiative transfer, and introduces minor modifications motivated by computational efficiency \citep{Zier2026}. Companion studies using \lumina have analyzed fluctuations in the cosmic microwave background optical depth \citep{Smith2026}, the demographics of high-redshift active galactic nuclei (AGNs), spanning luminous quasars and \textit{JWST} little red dots at $z\geq3$ \citep{Shen2026AGN}, and the inhomogeneous recombination budget and clumping evolution during reionization \citep{Sadain2026}.

Radiation transport is solved directly on the moving Voronoi mesh with \areport, coupled to a non-equilibrium primordial thermochemistry network for hydrogen and helium \citep{Kannan2019,zier2024adapting}. To reduce the radiative-transfer time-step constraint, the solver employs a reduced-speed-of-light approximation with $\tilde{c}=0.2\,c$ \citep{Gnedin2001,Kannan2022a}. During hydrogen reionization ($z>4.75$), the radiation field is discretized into six energy bins: three spanning the \HI, \HeI, and \HeII ionization thresholds, and three higher-energy X-ray bins. The dominant ionizing sources are stellar populations based on \textsc{bpass} v2.2.1 spectra \citep{BPASS2017,Stanway2018} and a Chabrier initial mass function \citep[IMF;][]{Chabrier2003}, with a constant stellar escape fraction $f_{\text{esc},\star} = 0.18$ calibrated to reproduce the desired hydrogen-reionization history. \lumina also includes hard radiation from accreting black holes and X-ray emission associated with high-mass X-ray binaries and hot ISM gas \citep{mcquinn2009he,Shen2020,2013Fragos-HMXB,2016Fragos-Erratum,Madau2017,2012bMineo-HotISM,2014Pacucci}. At $z=4.75$, once hydrogen reionization is effectively complete, the two lowest-energy photon bins are replaced by a spatially uniform metagalactic background, and the four remaining transported bins are merged to optimize the subsequent \HeII calculation \citep{Zier2026}.

\subsection{High-cadence Cartesian analysis products}

We use the Cartesian outputs of \lumina, which are written at intervals of approximately $2$--$4$\,Myr. Although primary fields are also available on a $2560^3$ grid, we analyze all fields on the common $1280^3$ uniform grid, which has a fixed cell size of $390\,\text{ckpc}$ \citep{Zier2026}. The deposition uses an adaptive second-order cloud-in-cell procedure in which each Voronoi cell is approximated as a cube and its mass and/or volume is distributed to neighboring voxels in proportion to geometric overlap. These outputs retain the three-dimensional structure of the ionization field at the analysis-grid resolution at each snapshot and are therefore the natural analysis product for identifying connected \HII regions and characterizing their internal properties. From these grids we use the volume-weighted hydrogen ionized fraction together with the corresponding density and radiation-energy fields. The analysis-grid resolution sets the minimum resolved bubble scale and affects the abundance of small regions, while the high cadence follows bubble growth and percolation through the full EoR. Sizes well above the grid scale are expected to be less sensitive to resolution, although we do not establish grid convergence here.

\subsection{Watershed identification of ionized bubbles}
\label{subsec:ws}

We identify ionized bubbles on the Cartesian grids using a three-dimensional watershed segmentation algorithm motivated by the analysis of \citet{Lin2016}, who showed that the watershed and mean-free-path estimators provide meaningful but distinct measures of bubble sizes in reionization simulations. The same estimator distinction was recently emphasized by \citet{Lu2024} in a galaxy-focused bubble analysis. We work on the volume-weighted $x_\text{\HII}$ field with periodic boundary conditions and define ionized cells as those with $x_\text{\HII}>0.5$.

We first compute the Euclidean distance transform,
\begin{equation}
d(\bm{r}) \equiv \min_{\bm{r}_\text{n}} \|\bm{r}-\bm{r}_\text{n}\| \, ,
\end{equation}
where $\bm{r}_\text{n}$ denotes the position of the nearest neutral cell with $x_\text{\HII}<0.5$. This is efficiently calculated for all cells using a pre-sorted distance stencil, accounting for periodicity and terminating early when the first neutral cell is found. Ionized cells deep inside a bubble, therefore, have large values of $d$, while cells close to an ionization front have small values. Local maxima of this distance field serve as the initial bubble seeds and thereby establish the initial number of watershed basins before the merging step described below. Specifically, we identify these maxima relative to their 26 adjacent neighbors, and in cases of exact degeneracy, a single cell is arbitrarily selected to represent the local maximum.

Starting from these seeds, we grow individual regions through the distance field until every ionized cell is assigned to exactly one watershed basin. Although the standard watershed algorithm is usually formulated by flooding upward from local minima, our implementation is equivalent to applying a conventional watershed to the negated distance field, so that individual bubbles are seeded at the deepest points inside the ionized regions. The segmented volumes expand iteratively from each seed to their sequentially highest-valued neighbors until all ionized cells are uniquely assigned, following a procedure similar to that described by \citet{Jamieson2025}, except that bubbles remain separate at this stage. When multiple growing bubbles compete for the same boundary cell, we resolve the conflict by assigning the cell to the progenitor bubble with the largest accumulated volume.

Watershed segmentation can split large connected structures into many small basins. A common image-processing remedy is an H-minima transform, which removes local minima shallower than a chosen contrast depth before watershed flooding. We instead suppress over-segmentation with a post-processing step based on the shared surface area. After the initial segmentation is complete, we merge neighboring basins whose shared interface area is at least $10\%$ of the total surface area of the smaller basin. In the idealized case of two merging spheres, this threshold corresponds to a solid angle overlap of $4\pi/10$ steradians for the smaller bubble, equivalent to a circular cap with half-opening angle $\theta \approx 37^\circ$.

\begin{figure*}
    \centering
    \includegraphics[width=\textwidth]{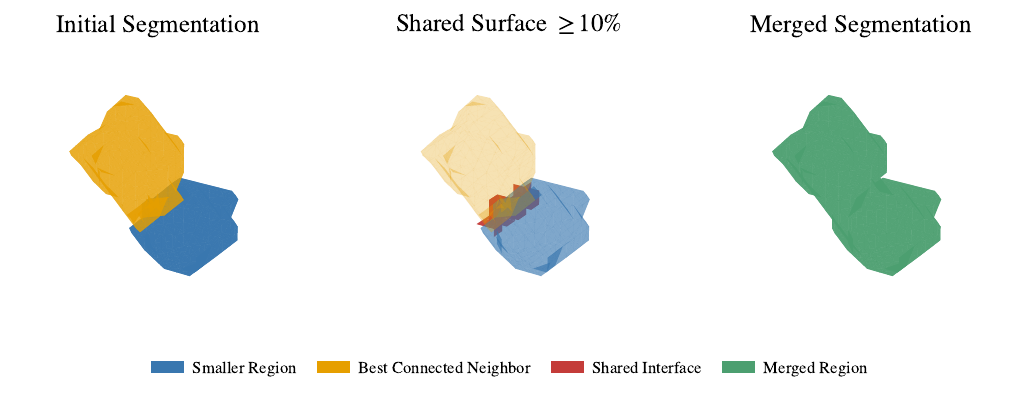}
    \caption[Watershed de-fragmentation example]{An example of watershed de-fragmentation in the $z=9$ \lumina ionization field. Left: the smaller watershed region (blue) and its best-connected neighboring region (yellow) before post-processing. Center: their shared grid-face interface (red). The two regions share 26 faces, while the smaller region has 200 total surface faces, giving $A_{\rm shared}/A_{\rm small}=26/200=0.13$. Because this exceeds the fiducial threshold of $10\%$, the regions are combined into the single green region shown at right. All three panels show the same snapshot, illustrating how the shared-surface criterion combines adjacent watershed basins.}
    \label{fig:ws_defragmentation}
\end{figure*}

Figure~\ref{fig:ws_defragmentation} illustrates this decision for one near-threshold pair in the $z=9$ segmentation. The denominator includes the complete grid-face surface of the smaller basin, including interfaces with the neutral background and with all neighboring watershed regions, rather than only the contact area of the displayed pair. We investigate the impact of this choice in Appendix~\ref{appx:threshold}, where we repeat the size and shape measurements for shared-surface thresholds between $2.5\%$ and $30\%$. The post-processing reduces fragmentation within large connected \HII complexes while retaining boundaries between more weakly connected regions.

\begin{figure*}
    \centering
    \includegraphics[width=\linewidth]{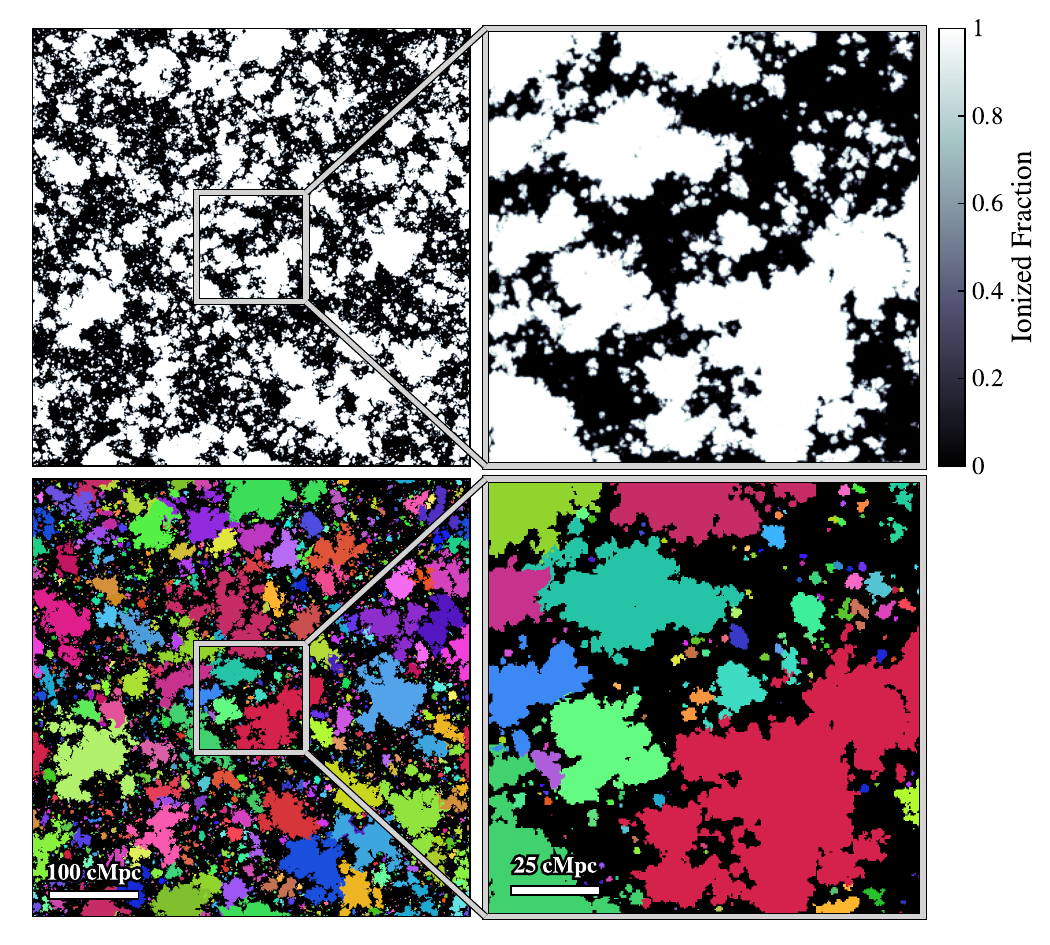}
    \caption[Bubble segmentation map]{A one-voxel-thick slice ($0.39$\,cMpc) through the \lumina simulation at a global volume-weighted neutral hydrogen fraction of $x_\HI=0.5$ ($z=6.9$) showing the volume-averaged ionized fraction (upper panels) and the unique bubbles in different colors as identified by the watershed algorithm (lower panels). The left panels show the full $500$\,cMpc box and the right panels show a zoomed-in region. Patches that appear disconnected in this slice can belong to the same bubble if they are connected outside the slice. Black regions in the lower panels are neutral cells not assigned to ionized bubbles.}
    \label{fig:segmentation}
\end{figure*}

The final output assigns a unique identifier to each ionized cell within a snapshot, thereby providing a complete catalog of the bubble population at each redshift. Figure~\ref{fig:segmentation} illustrates this segmentation halfway through reionization ($x_\text{\HII} = 0.5$ at $z = 6.9$). The upper panels show the volume-weighted ionized hydrogen fraction, while the lower panels show the bubbles identified in a slice through the full box (left) and a zoomed-in region (right).

At late times, when the ionized phase percolates across the box, the same framework can be applied to the residual neutral component by interchanging the roles of ionized and neutral cells. We use this inverted segmentation to characterize the neutral-island population in the final stages of hydrogen reionization.

\subsection{Derived bubble statistics}

For each segmented region $b$, we compute its volume
\begin{equation}
  V_b = N_b\,V_\text{cell} \, ,
\end{equation}
where $N_b$ is the number of grid cells assigned to the bubble and $V_\text{cell}$ is the Cartesian cell volume. We then define the effective spherical radius
\begin{equation} \label{eq:Reff}
  R_\text{eff} \equiv \left(\frac{3V_b}{4\pi}\right)^{1/3} \, .
\end{equation}
This quantity provides a compact scalar measure of the connected watershed volume, following the convention used in previous bubble-size studies \citep{Lin2016, Lu2024}. The full segmentation is retained for more general morphological analyses. We use the distance-transform field to define a second internal length scale,
\begin{equation}
  R_{{\rm DT},b} \equiv \frac{1}{N_b}\sum_{n\in b} d_n \, ,
  \label{eq:RDT}
\end{equation}
where $d_n$ is the shortest Euclidean distance from cell $n$ to a neutral cell. Thus, $R_{{\rm DT},b}$ characterizes the typical depth of the bubble interior below an ionization front. We also record the spatial coordinates of the initial local maxima, which act as the characteristic seeding nodes for their respective bubbles.

The segmentation allows us to compute bubble-averaged internal properties directly from the same grid cells. For any gridded field $q$, we may evaluate
\begin{equation}
  \bar{q}_b = \frac{1}{N_b}\sum_{n\in b} q_n \, ,
\end{equation}
which we use below for quantities such as the gas overdensity and the internal radiation-energy density. Because the bubble boundaries are defined in three dimensions, these averages characterize the actual connected ionized regions rather than one-dimensional lines of sight through them.

We similarly compute the geometric center of each bubble by assigning unit weight to every grid cell and evaluating the corresponding centroid, $\bar{\bm{r}}_b$, carefully accounting for periodic wrapping with respect to the seed position, $\bm{r}_{b,0}$. The watershed regions are not assumed to be ellipsoids: they can be non-convex, percolating, and multiply connected, especially near the end of reionization. The following shape tensor is therefore a second-moment descriptor of the full segmented volume, and terms such as ``oblate'' and ``prolate'' should be interpreted as shorthand for flattened and elongated second-moment structure. To quantify these shapes, we construct the geometric shape tensor
\begin{equation}
  \bm{\mathsf{S}} = \sum_{n\in b} (\bm{r}_n - \bar{\bm{r}}_b) \otimes (\bm{r}_n - \bar{\bm{r}}_b) \, ,
\end{equation}
where the sum runs over all cells in the bubble. Ordering the eigenvalues as $\lambda_0 \le \lambda_1 \le \lambda_2$, which scale as the squared principal semi-axes of the equivalent uniform ellipsoid, we quantify the morphological properties of segmented regions by defining the triaxiality parameter
\begin{equation}
  T = \frac{\lambda_2-\lambda_1}{\lambda_2-\lambda_0} \, .
  \label{eq:triax}
\end{equation}
Following the standard triaxiality convention of \citet{Franx1991}, we classify the bubble geometries as oblate for $T < 1/3$, triaxial for $1/3 < T < 2/3$, and prolate for $T > 2/3$. In this convention, oblate shapes have one minor and two major axes, giving them a flattened, pancake-like morphology. Prolate shapes instead have one major and two minor axes and are therefore elongated or filamentary, while triaxial geometries have three distinct principal axis lengths. Triaxiality distinguishes these shape classes but does not by itself measure the degree of departure from sphericity.

For direct comparison with earlier work, we compare the watershed effective radii with mean-free-path \citep{Mesinger2007} bubble sizes measured from the same ionization field using the second-order ray-tracing framework of \citet{Neyer2024}. For every segmented bubble, $R_{\rm MFP}$ is the cell-averaged one-sided ray distance to a neutral boundary, while $R_{\rm DT}$ is the cell-averaged shortest distance in any direction (Eq.~\ref{eq:RDT}). The watershed estimator $R_{\rm WS}\equiv R_{\rm eff}$ measures the connected three-dimensional volume. Both boundary-distance estimators terminate at neutral cells and can cross internal interfaces between adjacent watershed basins. The three quantities thus describe complementary aspects of bubble geometry, and differences between them are physically expected, particularly once bubbles become strongly non-spherical and highly connected \citep{Lin2016,Lu2024,Neyer2024}.

We distinguish full-population statistics, which describe bubble counts and the distribution of ionized volume, from conditional trends in median properties at fixed radius. Because large bubbles are rare, we typically display the radius-conditioned trends only for $R_\text{eff}\leq50$\,cMpc (Figures~\ref{fig:mfp_ws}, \ref{fig:triax_R}, \ref{fig:overdens_Reff}, and \ref{fig:ione_Reff}). This is a display convention rather than a measured convergence threshold, and the statistical support still depends on the full catalog data. Full-population distributions, counts, volumes, and size evolution include all structures, as do the joint distributions subject to the sample definitions in their captions.

\section{Ionized Bubble Sizes}
\label{sec:sizes}

\begin{figure}
    \centering
    \includegraphics[width=\myfigwidth]{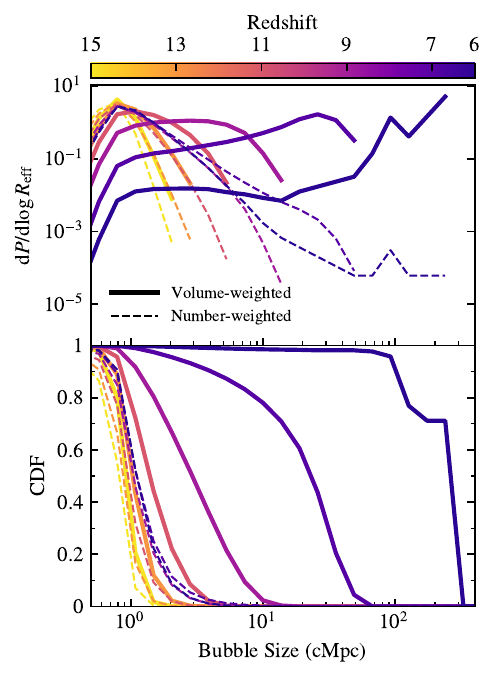}
    \caption[Lumina bubble size distributions]{Bubble size probability density functions (PDFs; top) and cumulative distributions (bottom, labeled CDF) throughout the EoR with both volume-weighted and number-weighted statistics. The lower panel shows the fraction above a given bubble size: the fraction of ionized volume for volume weighting, or of bubbles for number weighting. Bubble size is defined as the spherical effective radius ($R_\text{eff}$) corresponding to the watershed volume. The number-weighted curves are the standard object-by-object PDFs and therefore emphasize the numerous small, resolution-sensitive bubbles, while the volume-weighted curves show where the ionized volume resides. The main trend is the transfer of ionized volume from $\sim$1\,cMpc bubbles at early times to $\gtrsim100$\,cMpc connected structures by $z\approx6$.}
    \label{fig:bsd}
\end{figure}

\begin{figure}
    \centering
    \includegraphics[width=\myfigwidth]{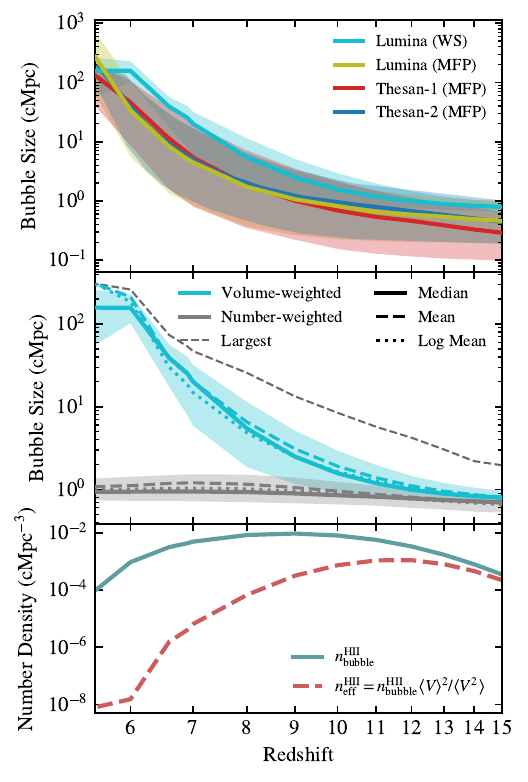}
    \caption[Bubble size evolution]{The upper panel shows median and $16^\text{th}$--$84^\text{th}$ percentile bubble size evolution comparing the watershed effective bubble radius in \lumina (cyan) to MFP bubble sizes in \lumina (yellow) and in \thesanone (red) and \thesantwo (blue) from \citet{Neyer2024}. All statistics are volume-weighted. The center panel shows \lumina bubble sizes according to the watershed method, with volume-weighted (cyan) and number-weighted (gray) statistics. The lower panel shows the raw bubble number density, $n_{\rm bubble}^{\HII} = N_{\rm bubble}^{\HII} / V_{\rm box}$, and the effective number density, $n_{\rm eff}^{\HII} = N_{\rm eff}^{\HII} / V_{\rm box} = n_{\rm bubble}^{\HII} \langle V\rangle^2 / \langle V^2\rangle$. The volume moments are number-weighted over the same complete sample of segmented bubbles. Both densities are in comoving units of cMpc$^{-3}$.}
    \label{fig:evol}
\end{figure}

Figure~\ref{fig:bsd} shows the bubble size distributions (BSDs) at several redshifts, presenting both the probability density and the cumulative fraction (CDF) above a given radius with volume-weighted and number-weighted statistics. Bubble size is defined as the spherical effective radius ($R_\text{eff}$ from Eq.~\ref{eq:Reff}) corresponding to the total segmented volume of each ionized region. The number-weighted distributions describe the population on an object-by-object basis, whereas the volume-weighted distributions identify the bubble scales in which most of the ionized volume resides. The volume-weighted PDFs evolve substantially across the EoR: at $z \gtrsim 10$, the distribution is sharply peaked near $R_\text{eff} \approx 1$\,cMpc and falls off steeply above a few cMpc, whereas by $z \approx 6$, most of the ionized volume lies in bubbles with $R_\text{eff} \gtrsim 100$\,cMpc and the cumulative curve falls almost as a step function at the largest resolved scales. The early $\sim$1\,cMpc peak should be interpreted with the analysis-grid scale in mind because the Cartesian voxel size is $0.39$\,cMpc, meaning that the corresponding radii span only a few voxel widths and their exact location is resolution dependent. Both the characteristic volume-weighted bubble size and the maximum bubble radius grow monotonically as redshift decreases. In contrast, the number-weighted distributions remain dominated by small bubbles ($R_\text{eff} \approx 1$\,cMpc) at all redshifts, reflecting the persistent abundance of small, recently nucleated bubbles throughout the EoR. Although the peak of the number-weighted PDF is nearly fixed in $R_\text{eff}$, it traces the smallest resolved source-centered structures and should not be interpreted as a converged physical lower cutoff. The redshift evolution in these number-weighted statistics appears mainly in the high-$R_\text{eff}$ tails, which extend to progressively larger radii at lower $z$. By contrast, the volume-weighted distributions show much stronger growth in the typical size of the dominant ionized regions as large bubbles coalesce and encompass an increasing fraction of the ionized volume.

We also examine the evolution of characteristic bubble scales throughout the EoR. The top panel of Figure~\ref{fig:evol} shows the median and $16^\text{th}$--$84^\text{th}$ percentile bubble radii as a function of redshift, comparing watershed sizes in \lumina (cyan) with MFP sizes in \lumina (yellow) and with the MFP sizes measured by \citet{Neyer2024} in \thesanone (red) and \thesantwo (blue). This comparison is useful because it follows the growth of a connected three-dimensional volume and the evolution of a characteristic line-of-sight path length in the same ionization field. For computational efficiency, the \lumina MFP bubble size at $z=5.5$ was calculated on a grid with 640 cells per side. This lower resolution is expected to have little effect when the IGM is almost entirely ionized and the bubbles are large, although a matched-resolution test would be needed to quantify its impact. At the earliest redshifts ($z \gtrsim 13$), the watershed curve flattens at $R_\text{eff} \approx 0.8$--$0.9$\,cMpc, whereas the MFP curve lies at approximately $0.3$--$0.5$\,cMpc. Both scales are comparable to the $0.39$\,cMpc voxel size and are therefore resolution dependent, although they still identify the smallest resolved, localized ionized regions around early sources. As reionization proceeds, the watershed-derived effective radii grow substantially faster than the MFP sizes, and by intermediate redshifts ($z \approx 7$--$9$), the offset reaches a factor of a few. The characteristic line-of-sight distance therefore grows more slowly than the volume-equivalent bubble radius. The matched watershed measurement in the larger \lumina volume extends the earlier \thesan MFP evolution from \citet{Neyer2024} and complements the \thesan bubble merger histories of \citet{Jamieson2025}, who found that rapid merging and the emergence of a dominant bubble drive late-time bubble growth. All curves in the upper panel reflect volume-weighted statistics.

The middle panel compares volume-weighted and number-weighted watershed radii in \lumina. The volume-weighted median radius grows from $\sim$1\,cMpc at high redshift to $\gtrsim100$\,cMpc near the end of reionization, corresponding to an increase in effective volume from a few $\text{cMpc}^3$ at $z = 15$ to $V \sim 10^7\,\text{cMpc}^3$ at $z \approx 6$. The number-weighted median radius remains close to $\sim$1\,cMpc throughout because the population by count is dominated by small bubbles near the resolution scale. The bottom panel shows the number density of watershed bubbles in \lumina. The initial number of watershed basins is set by the local maxima in the distance-transform field, corresponding to cells that are locally farthest from the nearest neutral boundary. The shared-surface merging step then reduces this number to the final $N_{\rm bubble}^{\HII}$. For the same sample of $N_{\rm bubble}^{\HII}$ bubbles in the comoving volume $V_{\rm box}$, we define
\begin{equation}
  N_{\rm eff}^{\HII} \equiv N_{\rm bubble}^{\HII} \frac{\langle V\rangle^2}{\langle V^2\rangle}
  = \frac{(\sum_i V_i)^2}{\sum_i V_i^2} \, ,
  \quad
  n_{\rm eff}^{\HII} = \frac{N_{\rm eff}^{\HII}}{V_{\rm box}} \, ,
  \label{eq:neff}
\end{equation}
where $\langle V^k \rangle = \sum_i V_i^k / N_{\rm bubble}^{\HII}$ denotes a number-weighted volume moment. Thus $N_{\rm eff}^{\HII}$ is dimensionless, whereas $n_{\rm eff}^{\HII}$ has units of cMpc$^{-3}$. For equal-volume bubbles, $N_{\rm eff}^{\HII}=N_{\rm bubble}^{\HII}$, while $N_{\rm eff}^{\HII}$ approaches unity when one bubble dominates the ionized volume. The effective number therefore responds to both the raw abundance and the inequality of bubble volumes. It describes the global population, while statistical support for conditional trends depends on the occupancy of individual bins. At early redshifts ($z \gtrsim 10$), the bubble number density increases as reionization accelerates and new luminous sources generate isolated \HII regions, reaching a broad maximum near $n_\text{bubble}^{\HII} \sim 10^{-2}\,\text{cMpc}^{-3}$ around $z \approx 9$. The effective number density remains far below the raw count density because only a subset of the identified bubbles carries most of the ionized volume, and it declines once percolation concentrates that volume into a small number of dominant structures. During the same phase, the raw number density falls as smaller bubbles merge into larger ionized structures, reaching $n_\text{bubble}^{\HII} \approx 10^{-3}\,\text{cMpc}^{-3}$ by $z \approx 6$ and $\lesssim 10^{-4}\,\text{cMpc}^{-3}$ by $z \approx 5.5$. This late-time behavior is consistent with the three-stage picture of \citet{Jamieson2025}, in which initial expansion gives way to rapid mergers and then to the accelerated growth of a dominant ionized structure.

\begin{figure}
    \centering
    \includegraphics[width=\myfigwidth]{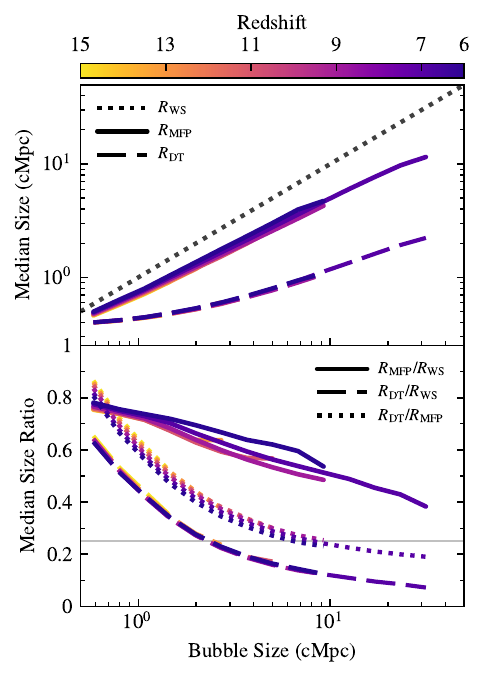}
    \caption[Comparison of bubble length scales]{Bubble length scales in \lumina. The top panel compares the median watershed volume-equivalent radius ($R_{\rm WS}$), bubble-averaged mean-free-path scale ($R_{\rm MFP}$), and mean distance-transform scale ($R_{\rm DT}$). The bottom panel shows their three pairwise comparisons, calculated as medians of per-object ratios. The shared horizontal coordinate is the watershed radius, $R_{\rm WS}=R_\text{eff}$. Colors denote redshift, and the silver line marks the uniform-sphere expectation $R_{\rm DT}/R_{\rm WS}=1/4$. Curves are shown over $0.5\leq R_{\rm WS}\leq50$\,cMpc, requiring at least 20 bubbles per estimator bin.}
    \label{fig:mfp_ws}
\end{figure}

Figure~\ref{fig:mfp_ws} compares the absolute scales and their ratios over the displayed $0.5$--$50$\,cMpc watershed-radius range. The medians follow the clear ordering $R_{\rm WS}>R_{\rm MFP}>R_{\rm DT}$. This ordering reflects what each estimator measures. The watershed scale converts the segmented volume to an equivalent spherical radius, the MFP scale averages one-sided ray distances, and the distance transform selects the shortest distance from each cell to neutral gas. Accordingly, $R_{\rm DT}$ is especially sensitive to nearby walls, narrow necks, and internal neutral structure, whereas $R_{\rm WS}$ continues to grow as distant branches join the same connected region.

The spherical reference values follow directly from volume averaging. For a sphere of radius $R$, a uniformly sampled point has $\langle r\rangle=3R/4$, so its mean shortest distance to the surface is $\langle R-r\rangle=R/4$ and hence $R_{\rm DT}/R_{\rm WS}=1/4$. The factor $3/4$ for the MFP scale has a different geometric origin. Fixing the ray direction by spherical symmetry, a ray at impact parameter $\rho$ lies on a chord with half-length $h=(R^2-\rho^2)^{1/2}$. Averaging the one-sided remaining distance $h-z$ over all uniformly distributed starting points, with longer chords receiving their proper volume weight, gives $\langle h-z\rangle=3R/4$ \citep{Smith2025}. Therefore a uniform sphere has $R_{\rm MFP}/R_{\rm WS}=3/4$ and $R_{\rm DT}/R_{\rm MFP}=1/3$.

At the smallest resolved scales, $R_{\rm MFP}/R_{\rm WS}$ lies close to its spherical value, while the distance-transform ratios remain affected by the finite voxel size. All three ratios then decline systematically as $R_{\rm WS}$ increases, with $R_{\rm DT}/R_{\rm WS}$ falling most rapidly. This increasing departure from the spherical limits reflects the branching and percolation of large connected regions, whose total volume continues to grow even though a typical interior cell remains comparatively close to an ionization front. The similar trends across redshift show that this scale dependence persists across the sampled epochs and reinforce that bubble size is not a unique, method-independent quantity. The watershed estimator provides a natural descriptor of the full three-dimensional topology relevant to tomographic 21\,cm studies \citep{Mesinger2007,Lin2016,Giri2018}, while MFP and distance-to-boundary scales are more closely connected to the path lengths relevant for Ly$\alpha$ damping-wing and transmission analyses \citep{Smith2022,Xu2023,Keating2023a,Mason2026,Neyer2026}.

\section{Ionized Bubble Shapes}
\label{sec:shapes}

\begin{figure}
    \centering
    \includegraphics[width=\myfigwidth]{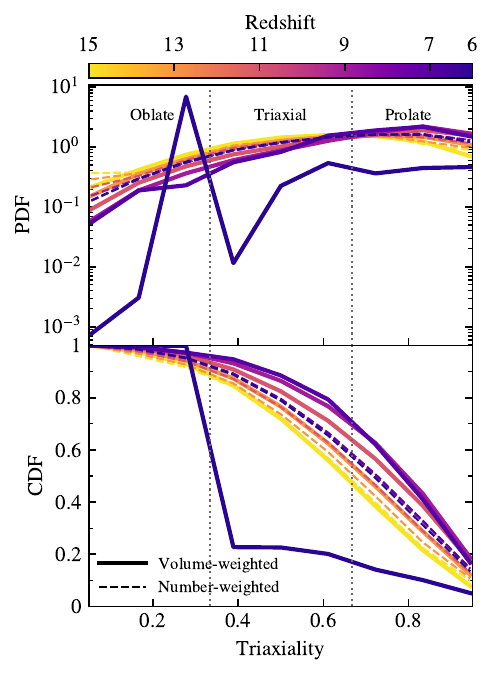}
    \caption[Ionized bubble shape distributions]{Probability density functions (top) and cumulative fractions above the indicated triaxiality (CDFs, bottom) during reionization, with volume-weighted curves shown as solid lines and number-weighted curves shown as dashed lines. Through most of the EoR, the distributions peak in the elongated/prolate regime. By $z \approx 6$, the volume-weighted distribution develops a distinct bimodal structure, with a sharp flattened/oblate peak near $T \approx 0.25$ alongside the persistent prolate population. Because this feature is absent from the number-weighted distribution, it reflects a small number of very large, volume-dominant structures rather than a generic property of the bubble population.}
    \label{fig:triax}
\end{figure}

\begin{figure}
    \centering
    \includegraphics[width=\myfigwidth]{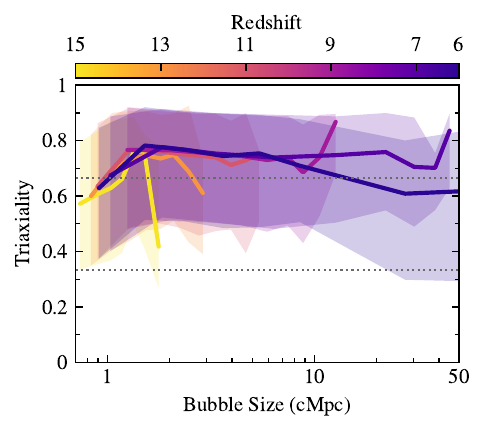}
    \caption[Shapes of bubbles according to their sizes]{Median triaxiality and $16^\text{th}$--$84^\text{th}$ percentile ranges as a function of bubble size, displayed over $0.7\leq R_\text{eff}\leq50$\,cMpc. Small and intermediate bubbles are predominantly prolate through most of the EoR. Adaptive bins include all their members, and the upper display limit reflects the sparse sampling of large bubbles.}
    \label{fig:triax_R}
\end{figure}

In addition to measuring bubble sizes, the watershed catalog allows us to characterize the three-dimensional morphology of individual ionized regions. As described in Section~\ref{sec:methods}, we use the eigenvalues of the shape tensor to compute the triaxiality parameter, $T$ (Eq.~\ref{eq:triax}), for which $T < 1/3$ indicates flattened second-moment structure, $1/3 < T < 2/3$ indicates triaxial structure, and $T > 2/3$ indicates elongated structure. Figure~\ref{fig:triax} shows the resulting triaxiality distributions across a range of redshifts. The number-weighted PDFs (dashed) peak in the prolate regime at all redshifts, with little evolution beyond modest changes in the height and width of the distribution. The volume-weighted PDFs (solid) vary more strongly with redshift: the highest-$z$ distributions are comparatively broad, while at intermediate redshifts ($z \approx 7$--$11$) the peak sharpens and shifts toward $T \approx 0.8$.

Observations of the galaxy--IGM transmission cross-correlation also provide evidence for anisotropic ionization geometry \citep{Zhu2026}. Near the end of reionization, the volume-weighted distribution at $z \approx 6$ changes qualitatively. A sharp peak appears at $T \approx 0.25$, implying that a substantial fraction of the ionized volume lies in structures with oblate second moments, while the volume-weighted density in the prolate regime drops well below the corresponding number-weighted level. This bimodality is essentially absent from the number-weighted statistics, indicating that it is associated with a small number of very large bubbles and suggesting late-time flattening during percolation.

\begin{figure}
    \centering
    \includegraphics[width=\myfigwidth]{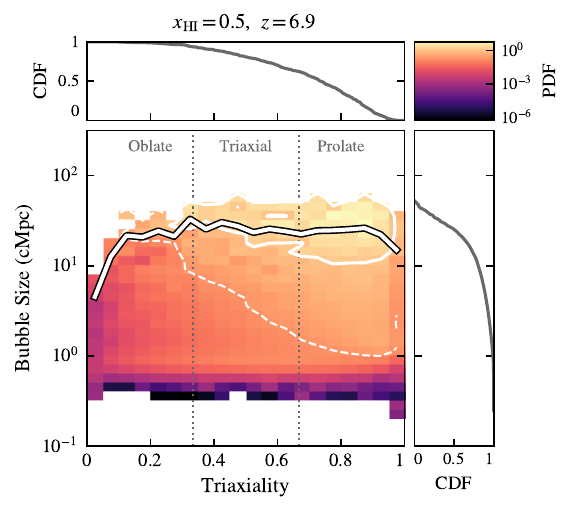}
    \caption[Joint bubble size and triaxiality]{Joint volume-weighted distribution of bubble triaxiality and effective radius $R_\text{eff}$ at the midpoint of reionization ($x_\text{H\,I}=0.5$, $z=6.9$). The sample includes bubbles with finite triaxiality and radius and positive finite volume, after excluding the single largest-volume bubble among these valid entries. All panels use this sample without an upper-radius selection. The color scale shows the volume-weighted PDF, the solid black-outlined curve traces the median bubble size in each triaxiality bin, and the white solid and dashed contours enclose the $1\sigma$ and $2\sigma$ regions, respectively. Marginal CDFs give the volume fraction above the indicated triaxiality (top) or size (right), and dotted vertical lines mark the oblate/triaxial ($T=1/3$) and triaxial/prolate ($T=2/3$) boundaries. Much of the ionized volume resides in prolate bubbles, while the most strongly oblate bin has a smaller characteristic radius than the triaxial and prolate population.}
    \label{fig:R_triax}
\end{figure}

To systematically explore these morphological trends, we examine bubble triaxiality as a function of effective radius. Figure~\ref{fig:triax_R} displays the median and $16^\text{th}$--$84^\text{th}$ percentile ranges over $0.7\leq R_\text{eff}\leq50$\,cMpc. For most of the EoR, the small and intermediate bubble populations remain predominantly prolate ($T\gtrsim0.7$). At high redshifts ($z\gtrsim13$), the relation turns toward more triaxial shapes at the upper end of the populated size range, whereas at intermediate redshifts ($7\lesssim z\lesssim9$) the medians are close to $T\approx0.75$ across small to medium radii. The populated radius range and bin occupancy vary with redshift, with sparse sampling at the largest radii. The full-population distribution in Figure~\ref{fig:triax} complements these conditional medians by showing the contribution of the largest structures to the total ionized volume.

The morphology of the volume-dominant ionized structures therefore changes qualitatively during late percolation. For most of the EoR, both the number- and volume-weighted distributions peak in the prolate regime, consistent with anisotropic growth through the cosmic web, whereas by $z \approx 6$, a strong oblate component appears in the volume-weighted distribution alone. The absence of a corresponding feature in the number-weighted distribution associates this component with a small number of exceptionally large structures, consistent with the late-time dominance of large connected regions in previous studies \citep{Giri2018, Jamieson2025}. We therefore interpret the late-time oblate peak as evidence for a morphological transition in the dominant connected \HII network in the fiducial segmentation, subject to the threshold sensitivity discussed in Appendix~\ref{appx:threshold}, while the number-weighted distribution continues to peak in the prolate regime.

Having examined the redshift evolution of the shape distribution and the dependence of median shape on radius, we now consider the joint distribution of shape and size at a representative stage of reionization. Figure~\ref{fig:R_triax} shows this volume-weighted joint distribution at the midpoint of reionization ($x_\HI=0.5$, $z=6.9$), when much of the ionized volume resides in prolate bubbles. The median bubble size rises from $R_\text{eff}\approx5$\,cMpc in the most strongly oblate bin to a broad plateau of $R_\text{eff}\approx20$--$30$\,cMpc across the triaxial and prolate regimes. The most strongly oblate bin has a smaller characteristic radius, but oblate structures span a broad range of sizes. The volume weighting emphasizes the contribution of the larger structures, consistent with anisotropic growth into elongated, filamentary regions.

\section{Ionized Bubble Interiors}
\label{sec:interiors}

\begin{figure}
    \centering
    \includegraphics[width=\myfigwidth]{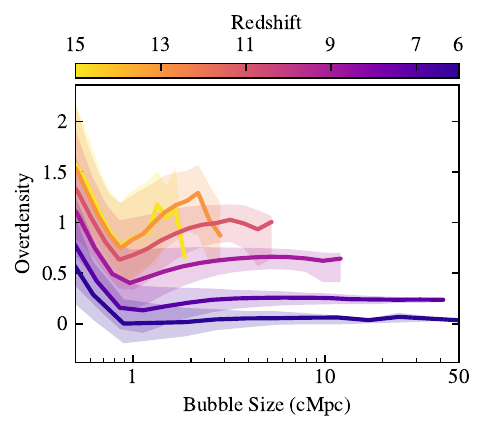}
    \caption[Overdensity within ionized bubbles]{Median and $16^\text{th}$--$84^\text{th}$ percentile ranges for gas overdensity, $\delta = (\rho - \bar{\rho})/\bar{\rho}$, in bubbles of different sizes, displayed over $0.5\leq R_\text{eff}\leq50$\,cMpc. At each redshift, the bubble-averaged gas overdensity in bubbles $\lesssim1$\,cMpc is elevated relative to the cosmic mean. At $z=15$, the smallest plotted bubbles have median $\delta\approx1.5$. The small-bubble median decreases to $\delta\approx0.2$--$0.3$ by $z=7$. Larger bubbles within the displayed range have lower median overdensities, closer to the cosmic mean.}
    \label{fig:overdens_Reff}
\end{figure}

\begin{figure}
    \centering
    \includegraphics[width=\myfigwidth]{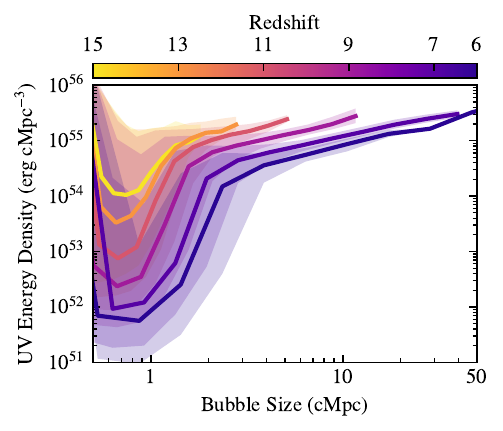}
    \caption[UV radiation energy density within bubbles]{Bubble-interior UV radiation energy density as a function of watershed bubble size, displayed over $0.5\leq R_\text{eff}\leq50$\,cMpc. For each bubble, the energy in the first three transported radiation bins ($13.6$--$100$\,eV) is summed over its cells and divided by its comoving volume. Solid curves show the number-weighted median across bubbles in each size bin, and shaded regions enclose the $16^{\rm th}$--$84^{\rm th}$ percentiles. Colors denote redshift. The median energy density rises with bubble size beyond a minimum close to the analysis-grid scale, where the result may be resolution dependent.}
    \label{fig:ione_Reff}
\end{figure}

We now relate bubble sizes and shapes to the physical state of the gas enclosed within each \HII region. Because the watershed segmentation explicitly defines the three-dimensional boundary of every bubble, we can compute the gas density and radiation-energy density over the full connected ionized volume. The present analysis focuses on the gridded gas and radiation fields, while the demographics of the galaxies enclosed by each bubble and the corresponding bubble-membership catalogs are deferred to future work.

Figure~\ref{fig:overdens_Reff} shows the bubble-averaged gas overdensity as a function of effective radius, $R_\text{eff}$. We average the local field $\delta = (\rho - \bar{\rho})/\bar{\rho}$ over each segmented volume, where $\bar{\rho}$ is the mean gas density in the simulation at that redshift. At very small bubble sizes ($R_\text{eff}\lesssim1$\,cMpc), the gas overdensity is consistently elevated above the cosmic mean, indicating that small bubbles form preferentially around overdense regions hosting the earliest luminous sources. The amplitude of this small-bubble overdensity evolves strongly with redshift. At $z=15$, the smallest plotted bubbles have a median overdensity of $\delta\approx1.5$, corresponding to a gas density about $2.5$ times the cosmic mean. The small-bubble median declines to $\delta\approx1$ by $z\approx11$ and to $\delta\approx0.2$--$0.3$ by $z=7$ as new bubbles increasingly form around more typical, less extreme overdensities. Larger bubbles have lower median overdensities, closer to the cosmic mean. This trend is consistent with ionization extending from dense source environments into surrounding lower-density gas.

\begin{figure}
    \centering
    \includegraphics[width=\myfigwidth]{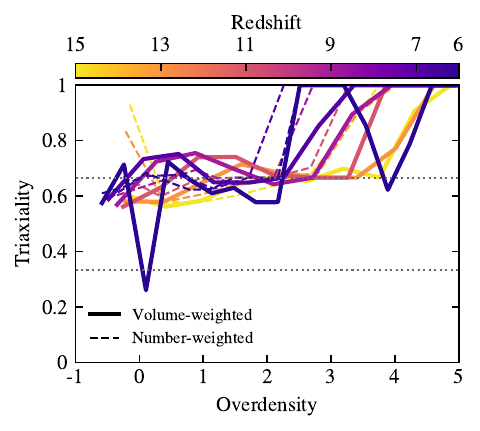}
    \caption[Triaxiality as a function of gas overdensity]{Volume- and number-weighted medians for triaxiality according to gas overdensity, $\delta = (\rho - \bar{\rho})/\bar{\rho}$. By number, the bubbles in the lowest and highest density regions are the most prolate, but by volume, the bubbles in the densest regions are preferentially more prolate than those in less dense regions, particularly before a few large bubbles dominate the statistics at the tail end of reionization. This trend is consistent with anisotropic growth in overdense environments.}
    \label{fig:triax_overdens}
\end{figure}

Figure~\ref{fig:ione_Reff} shows the mean UV radiation energy density within each bubble, using the first three radiation bins ($13.6$--$100$\,eV; \citealt{Zier2026}). For a bubble $b$ of comoving volume $V_b$, we calculate $\bar{u}_{\mathrm{UV},b}=V_b^{-1}\sum_{c\in b}\sum_{g=0}^{2}E_{g,c}$, where $E_{g,c}$ is the radiation energy in grid cell $c$ and frequency bin $g$. The resulting energy density is expressed in erg\,cMpc$^{-3}$. The plotted medians and percentile ranges give each bubble equal weight within its radius bin. Beyond a minimum near $R_\text{eff}\approx0.6$--$0.9$\,cMpc, the median UV energy density rises with bubble size over the plotted range. The minimum lies close to the analysis-grid scale, so its position and amplitude may be resolution dependent. At a fixed radius, the median also varies with redshift. These distributions describe the radiation field across the bubble population at each epoch. Connecting their evolution to individual sources, radiative transport, and physical mergers will require tracking regions between snapshots.

\begin{figure}
    \centering
    \includegraphics[width=\myfigwidth]{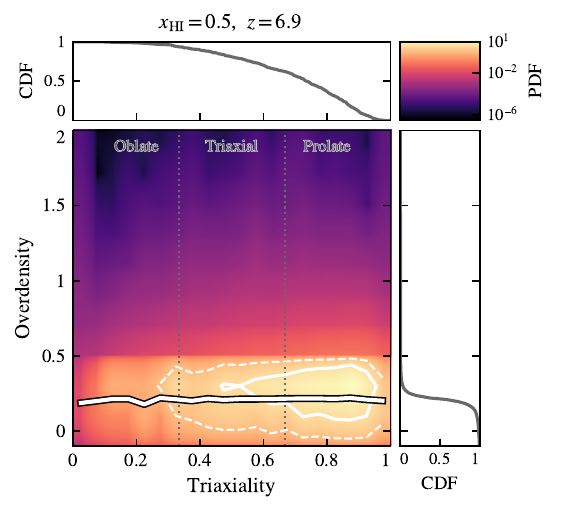}
    \caption[Joint gas overdensity and triaxiality]{As in Figure~\ref{fig:R_triax}, but for the joint volume-weighted distribution of triaxiality and internal gas overdensity $\delta = (\rho -\bar\rho)/\bar\rho$. The bulk of the ionized volume sits at modest overdensity ($\delta \approx 0.2$) and is predominantly prolate, with the median overdensity nearly flat across triaxiality. We retain bubbles with finite triaxiality and overdensity and positive finite volume, then exclude the single largest-volume bubble among these valid entries. The displayed range, $-0.1 \leq \delta \leq 2$, contains more than $99.5\%$ of the retained ionized volume. The PDF normalization, contours, median, and marginal CDFs (volume fractions above the indicated triaxiality or overdensity) use the full retained sample.}
    \label{fig:delta_triax}
\end{figure}

Having established how bubble interiors and their host environments evolve with bubble size and redshift, we next ask how bubble shape relates to the gas density of the region it occupies. Figure~\ref{fig:triax_overdens} shows the volume- and number-weighted median triaxiality as a function of internal gas overdensity $\delta$. By number, the relation is non-monotonic: bubbles in both the most underdense and most overdense regions are the most prolate, while bubbles at intermediate densities have lower median triaxiality. By volume, the densest regions preferentially host more prolate bubbles than less dense regions across most of the EoR, consistent with anisotropic growth of the volume-dominant population along overdense structures. This trend persists until the tail end of reionization, when a small number of very large bubbles begin to dominate the volume-weighted statistics and the relation becomes noisier. In conjunction with the monotonic decline of small-bubble overdensity in Figure~\ref{fig:overdens_Reff}, the density dependence supports an inside-out picture in which ionized regions form in overdense source environments and expand into the surrounding lower-density IGM.

The joint distribution at a single epoch complements the conditional trends in Figure~\ref{fig:triax_overdens}. Figure~\ref{fig:delta_triax} shows the volume-weighted joint distribution of triaxiality and internal gas overdensity at the midpoint of reionization ($x_\HI=0.5$, $z=6.9$). The volume-weighted population is concentrated at modest overdensities just above the cosmic mean ($\delta \approx 0.2$) and is predominantly prolate, consistent with the bulk population residing in mildly overdense, filamentary environments at this epoch. The median gas overdensity varies little with shape, differing by $\lesssim 0.1$ between the oblate and prolate regimes in this volume-weighted sample. A faint high-overdensity tail ($\delta \gtrsim 1$) extends across all triaxialities and is somewhat more populated toward the prolate end, consistent with the preference for prolate shapes in the densest environments. Together with Figures~\ref{fig:overdens_Reff} and \ref{fig:triax_overdens}, this snapshot connects the prevalence of prolate shapes to the mildly overdense interiors of the volume-dominant bubbles.

\section{Neutral Islands at Late Times}
\label{sec:neutral}

\begin{figure}
    \centering
    \includegraphics[width=\myfigwidth]{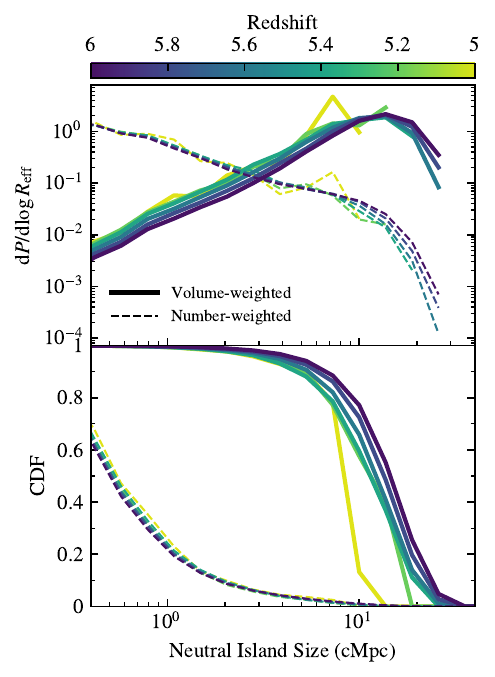}
    \caption[Neutral-island size distributions]{Neutral-island size PDFs (top) and cumulative fractions above the indicated size (CDFs, bottom) during the tail end of reionization ($5<z<6$), in the same format as Figure~\ref{fig:bsd}. The neutral islands shrink as the final pockets of the IGM become ionized. Their volume-weighted distribution is concentrated near $R_\text{eff}\sim10$\,cMpc between $z\approx6$ and $z\approx5.4$, whereas the number-weighted distribution is dominated by many small, resolution-sensitive islands.}
    \label{fig:neut_bsd}
\end{figure}

\begin{figure}
    \centering
    \includegraphics[width=\myfigwidth]{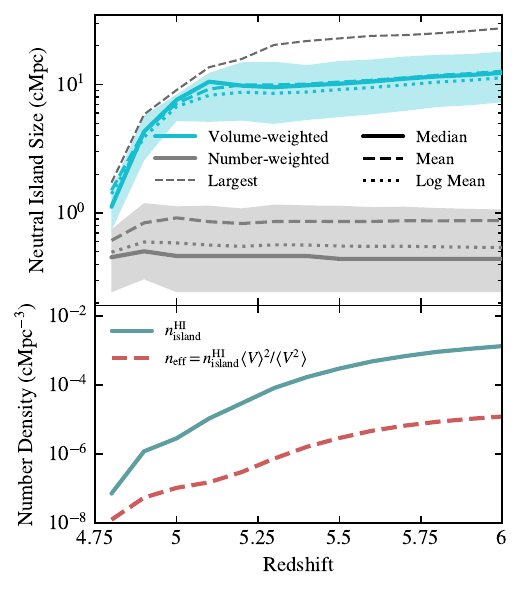}
    \caption[Neutral-island size evolution]{Evolution of neutral-island size and number density from the watershed catalog. The upper panel compares volume- and number-weighted median, mean, and logarithmic-mean radii, together with the largest island. Shaded regions show the $16^\text{th}$--$84^\text{th}$ percentile ranges. The lower panel shows the raw neutral-island number density, $n_{\rm island}^{\HI}$, and the effective number density $n_{\rm eff}=n_{\rm island}^{\HI}\langle V\rangle^2/\langle V^2\rangle$, analogous to Equation~\ref{eq:neff}. Both densities are in cMpc$^{-3}$. The volume-weighted island scale remains near $\sim$10\,cMpc through much of $5<z<6$, while the number density drops rapidly as the remaining neutral gas disappears. The plotted evolution extends to $z\approx4.8$ to show the disappearance of the final grid-scale islands just below the nominal analysis window.}
    \label{fig:neut_evol}
\end{figure}

\begin{figure}
    \centering
    \includegraphics[width=\myfigwidth]{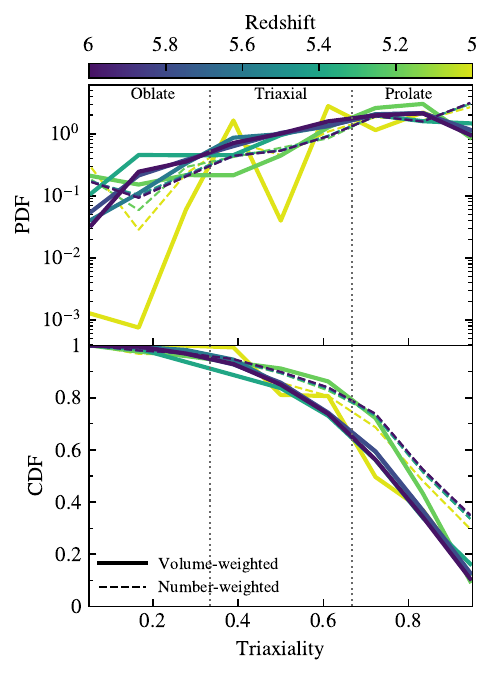}
    \caption[Neutral-island triaxiality distributions]{Neutral-island triaxiality PDFs (top) and cumulative fractions above the indicated triaxiality (CDFs, bottom) at $5<z<6$, when the IGM is mostly ionized and only residual neutral regions remain. Compare to the ionized-bubble triaxiality distributions in Figure~\ref{fig:triax}. Both number- and volume-weighted neutral-island statistics remain biased toward elongated/prolate morphologies, although the lowest-redshift curves are noisier because very few neutral islands remain.}
    \label{fig:neut_triax}
\end{figure}

After the ionized phase percolates, assigning most ionized gas to discrete bubbles becomes less informative because a single connected ionized network dominates the volume. The complementary late-time structures are the residual neutral islands embedded within that network. Persistent-homology analyses show that these islands detach late from the percolating neutral component and then disappear quickly once they are surrounded by ionized gas \citep{Elbers2023}. This topology motivates applying the watershed framework to the neutral component in \lumina during the tail end of reionization ($5<z<6$). We use the same shared-surface area merging procedure after interchanging the roles of ionized and neutral cells.

Figure~\ref{fig:neut_bsd} shows the probability density function of the effective radius of neutral islands. Between $z \approx 6$ and $z \approx 5.4$, the volume-weighted distribution peaks near $R_\text{eff} \approx 10$\,cMpc, with the largest individual islands extending to $R_\text{eff} \approx 20$--$30$\,cMpc. The number-weighted distributions instead peak at the resolution scale, indicating that the population by count is dominated by many small neutral patches even at these late times. Both distributions shift rapidly to smaller sizes at $z \lesssim 5.4$ as the final pockets of neutral IGM are ionized, and the volume-weighted PDF shifts toward the small-island tail. Figure~\ref{fig:neut_evol} shows the same evolution in terms of median, mean, and largest island radii: the volume-weighted median remains of order $10$\,cMpc across much of the late tail, while the number-weighted median stays near the smallest resolved islands. The raw neutral-island number density decreases steeply toward lower redshift as islands disappear, and $n_{\rm eff}$ shows that the remaining neutral volume is represented by only a small effective number of comparably important structures. The rapid contraction is qualitatively consistent with the low persistence of neutral islands found by \citet{Elbers2023}. The characteristic sizes are also in the regime targeted by recent photometric IGM-tomography forecasts, which show that late neutral islands spanning tens of cMpc can leave observable signatures in reconstructed Ly$\alpha$ forest transmission maps \citep{Giri2025}.

Figure~\ref{fig:neut_triax} shows that the distribution of neutral island shapes late in reionization is qualitatively similar to that of the ionized bubbles throughout the bulk of reionization. Both the volume- and number-weighted PDFs are biased toward the prolate regime, though with a broader distribution across triaxial values than the early-EoR ionized population. At the lowest redshifts in our sample ($z \lesssim 5.4$), the volume-weighted PDF becomes increasingly noisy as the total number of neutral patches drops and the statistics become dominated by a handful of remaining structures. The persistent prolate bias suggests that the last neutral regions remain anisotropic as they contract, plausibly reflecting residual neutral structure along the cosmic web. A direct environmental decomposition would be required to determine whether these neutral islands are preferentially associated with dense filaments, void boundaries, or self-shielded systems.

\section{Discussion}
\label{sec:discussion}

A central implication of our results is that the connection between simulations and observations depends on which aspect of bubble topology is being measured. Three-dimensional segmentations characterize the connected volume, shape, and internal environment of \HII regions, whereas MFP measurements sample distances to the first neutral boundary along rays, and damping-wing analyses constrain the distribution of neutral gas along source sightlines. A quoted ``bubble size'' can therefore refer to physically different geometric quantities even when the measurements are made from the same ionization field. Because the offset between these measures in \lumina is both strong and scale dependent, comparisons between simulations and observations should use matched estimators whenever possible \citep{Lin2016,Lu2024}. This requirement is especially relevant for recent \textit{JWST}-era proposals to map bubble edges and infer bubble sizes from Ly$\alpha$ transmission around ensembles of galaxies \citep{Lu2025,Nikolic2025}.

The larger \lumina volume is especially useful at late times, when rare connected structures carry most of the ionized volume. Earlier tests found little variation in the reionization history among eight sub-boxes of \thesanone \citep{Garaldi2022}. The larger \lumina calculation samples variation in reionization timing across a wider range of environments and improves the sampling of the late-time tail of the bubble distribution \citep{Zier2026, Smith2026}. In this regime, the watershed catalogs complement the \thesan MFP evolution of \citet{Neyer2024} and the merger-history analysis of \citet{Jamieson2025}. Future work will examine the evolution of individual watershed bubbles, their spatial correlations, and the hierarchical structure of the bubble size distribution.

The neutral-island extension places the segmentation framework in the context of residual neutral structure after percolation. Early work proposed searches for residual neutral islands in quasar spectra near $z \approx 5.5$ \citep{Malloy2015}, while the discovery of an extreme Ly$\alpha$ trough below $z=6$ established large opacity fluctuations near the end of reionization \citep{Becker2015}. Late-ending reionization simulations then showed that large \HI islands persisting below $z=5.5$ can generate strong Ly$\alpha$ opacity fluctuations and long troughs \citep{Kulkarni2019,Keating2020,Nasir2020}. Recent topology and tomography work provides a complementary view by emphasizing the short persistence of detached neutral islands and the observational value of mapping the rare survivors near the end of reionization \citep{Elbers2023,Giri2025}.

The observational context extends beyond trough length. Measurements of IGM opacity, transmission spikes, dark-gap statistics, and damping-wing-like absorption provide evidence for a late and spatially inhomogeneous tail of reionization at $z \lesssim 6$ \citep{Eilers2018,Yang2020,Zhu2021,Zhu2024,Spina2024,Becker2024}. The mapping from dark gaps to neutral islands is not one-to-one because low transmission can also arise in highly ionized gas \citep{Gnedin2022gaps}. Damping-wing profiles provide an additional test: fully coupled simulations can reproduce dark-gap statistics yet fail to reproduce the observed damping-wing-like features \citep{Gnedin2025}. The \lumina catalogs provide an estimator-aware, three-dimensional benchmark for different environments during the late stages of reionization, when percolation, non-sphericity, and rare connected structures increasingly shape observational outcomes. The measured distributions and connectivity of neutral islands are relevant for interpreting $z \gtrsim 5$ quasar spectra and galaxy--IGM connections \citep{Garaldi2025b,Garaldi2025a}. They are also relevant to 21\,cm--Ly$\alpha$ cross-correlation studies \citep[e.g.,][]{Chen2025}, in which neutral-region geometry affects the spatial relationship between 21\,cm emission and Ly$\alpha$ transmission.

As percolation concentrates the ionized volume into a few large structures, volume-weighted statistics become more sensitive to how those structures are segmented. The threshold sweep in Appendix~\ref{appx:threshold} shows increasing sensitivity of characteristic sizes toward the end of reionization (Figure~\ref{fig:threshold_evolution}), accompanied by a decline in the effective number of contributing bubbles (Figure~\ref{fig:threshold_neff}). Size and shape distributions remain qualitatively stable at the better-sampled epochs (Figure~\ref{fig:threshold_distributions}). At $z\lesssim7$, the partition of individual dominant regions has greater influence. The late-time oblate peak in the fiducial catalog (Figure~\ref{fig:triax}) therefore characterizes the morphology of these volume-dominant structures, with its detailed amplitude and shape subject to segmentation uncertainty.

\section{Conclusions}
\label{sec:conclusions}

We have characterized ionized bubbles in the \lumina simulation, which combines a large cosmological volume of $(500\,\text{cMpc})^3$ with resolved moving-mesh hydrodynamics, the IllustrisTNG galaxy-formation model, black-hole growth and feedback, non-equilibrium primordial thermochemistry, and on-the-fly radiation transport. The large volume allows the self-consistent galaxy-driven reionization calculation to sample rare sources, large-scale environments, cosmic variance, and the percolated structures that dominate late reionization. We identify discrete ionized bubbles with a three-dimensional watershed algorithm based on the distance transform from each cell to the nearest neutral cell, defined by $x_\text{\HII} < 0.5$. We then measure the sizes, shapes, and interiors of these bubbles throughout the Epoch of Reionization and apply the same framework to residual neutral islands near its end.

Our main findings are summarized as follows.
\begin{enumerate}
    \item The ionized bubbles in \lumina grow as reionization progresses, with the volume-weighted median bubble size growing from $R_\text{eff} \sim 1$\,cMpc at $z \gtrsim 10$ to $R_\text{eff} \sim 10$\,cMpc at $z \sim 7$--$8$. The end of the EoR is characterized primarily by very large bubbles, with $R_\text{eff} \gtrsim 100$\,cMpc at $z \lesssim 6$ (Figure~\ref{fig:evol}). The watershed-derived effective radii are systematically and substantially larger than the corresponding MFP measurements, by a factor of a few at intermediate redshifts. This offset is expected because the two estimators measure connected volume and line-of-sight boundary distances, respectively \citep{Lin2016,Lu2024}.

    \item The bubble number density rises as new sources form, peaks near $n_\text{bubble}^{\HII} \sim 10^{-2}\,\text{cMpc}^{-3}$ around $z \approx 9$, and declines as bubbles merge during the tail end of reionization ($z \lesssim 7$). At the same time, the effective number density falls as the ionized volume becomes concentrated in a few dominant structures (Figure~\ref{fig:evol}, lower panel). This evolution is consistent with the rapid merger phase and dominant-bubble growth described by \citet{Jamieson2025}. The absolute counts remain sensitive to small structures near the $0.39$\,cMpc analysis-grid scale corresponding to the $1280^3$ effective Cartesian resolution.

    \item The bulk bubble population remains predominantly prolate ($T \gtrsim 0.7$) across a wide range of effective radii for most of the EoR. The median triaxiality remains nearly constant with radius across the small and intermediate bubble populations, so the elongated morphology is not driven by size (Figure~\ref{fig:triax_R}).

    \item By $z\approx6$, the full-population volume-weighted triaxiality distribution develops a strong oblate component associated with a few large connected structures, while the number-weighted distribution remains peaked in the prolate regime (Figure~\ref{fig:triax}). This marks a change in the second-moment morphology of the volume-dominant ionized network in the fiducial segmentation. Its detailed form depends on how these structures are partitioned, as discussed in Appendix~\ref{appx:threshold}.

    \item Bubble interiors connect morphology to the evolving gas and radiation fields. Small bubbles occupy overdense environments, with their median overdensity declining over time, while larger bubbles have densities closer to the cosmic mean (Figure~\ref{fig:overdens_Reff}). Beyond a minimum near the analysis-grid scale, the median internal UV radiation energy density increases with radius within the displayed $0.5$--$50$\,cMpc range (Figure~\ref{fig:ione_Reff}).

    \item The same morphological framework provides a first \lumina characterization of the residual neutral medium at the close of reionization. Neutral islands at $5 \lesssim z \lesssim 6$ exhibit a volume-weighted size distribution that peaks near $R_\text{eff} \sim 10$\,cMpc, roughly an order of magnitude larger than the characteristic ionized bubble size at the start of the EoR ($R_\text{eff} \sim 1$\,cMpc at $z \gtrsim 10$). Like the early-EoR ionized bubbles, the neutral islands are predominantly prolate (Figure~\ref{fig:neut_triax}), suggesting that the last neutral structures remain anisotropic as they contract. The neutral-island size distribution collapses rapidly below $z \approx 5.4$, contracting from $\sim 10$-cMpc scales to the resolution limit over $\Delta z \lesssim 0.5$ (Figure~\ref{fig:neut_evol}). This rapid contraction is consistent with topology-based studies in which neutral islands detach late from the percolating neutral network and disappear quickly \citep{Elbers2023}. These late-time neutral structures are also relevant for tomography forecasts that target large residual islands near the end of reionization \citep{Giri2025}.
\end{enumerate}

By measuring the evolution and properties of bubbles as segmented volumes, these results complement the line-of-sight quantities commonly used in observational studies. This estimator-aware perspective clarifies how \lumina can be compared with current and upcoming constraints on bubble sizes, bubble environments, and late-time residual neutral structure.

\section*{Data Availability}
The \lumina simulation outputs are not currently publicly available. Requests for access, including requests for the derived catalogs and plotting data underlying this article, may be submitted through the project data-access page at \url{https://lumina-simulation.com/data-access/}.

\section*{Acknowledgments}
% We thank the referee for constructive comments and suggestions which have improved the quality of this work.
% We thank ... for insightful discussions related to this work.
An award of computer time was provided by the INCITE program. This research used resources of the Oak Ridge Leadership Computing Facility at the Oak Ridge National Laboratory, which is supported by the Advanced Scientific Computing Research programs in the Office of Science of the U.S. Department of Energy under Contract No.\ DE-AC05-00OR22725.
The authors acknowledge the MIT Office of Research Computing and Data and FAS Division of Science Research Computing Group at Harvard University for providing resources that have contributed to the research results reported within this paper.
Support for programs JWST-AR-08709 (AS) and JWST-AR-04814 (XS, MV) was provided by NASA through a grant from the Space Telescope Science Institute, which is operated by the Association of Universities for Research in Astronomy, Inc., under NASA contract NAS 5-03127.
Support for OZ was provided by Harvard University through the Institute for Theory and Computation Fellowship.
RK acknowledges support of the Natural Sciences and Engineering Research Council of Canada (NSERC) through a Discovery Grant and a Discovery Launch Supplement (funding reference numbers RGPIN-2024-06222 and DGECR-2024-00144) and York University's Global Research Excellence Initiative.
MV acknowledges support through NASA ATP Grant 23-ATP23-149 and NSF AAG Grant AST-2307699.
This work was supported by the LOEWE programme of the State of Hesse (LOEWE-Spitzen-Professur).
VS and LH acknowledge support from the Simons Foundation through the ``Learning the Universe'' initiative.
This work was supported in part by a grant of access to OpenAI models through the ChatGPT for Academic Researchers program.
Generative AI tools were used to assist with language editing and consistency checks. The authors remain responsible for the scientific content and final wording of the manuscript.

\bibliographystyle{mnras}
\bibliography{new_thesis}

\begin{appendix}
\renewcommand{\theHfigure}{A.\arabic{figure}}

\section{Threshold Sensitivity}
\label{appx:threshold}

The watershed post-processing described in Section~\ref{subsec:ws} merges a smaller basin into its most strongly connected neighbor when their shared interface accounts for at least a fraction $f_{\rm merge}$ of the smaller basin's total surface area. Our fiducial catalogs adopt $f_{\rm merge}=10\%$. To test the sensitivity of our conclusions to this choice, we repeat the merger step for $f_{\rm merge} \in \{2.5, 5, 10, 20, 30\}\,\%$, starting from the same unmerged watershed segmentation in each case.

\begin{figure}
    \centering
    \includegraphics[width=\myfigwidth]{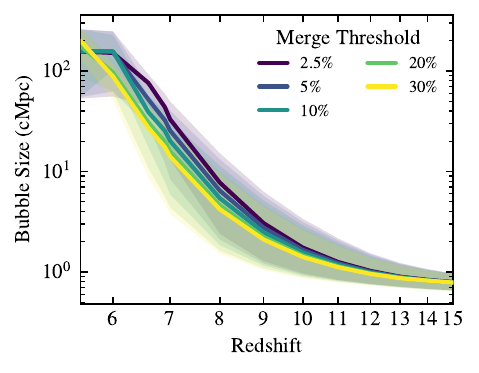}
    \caption[Bubble-size sensitivity to the merge threshold]{Evolution of the volume-weighted median watershed effective radius for five choices of the shared-surface merging threshold. Shaded regions enclose the $16^{\rm th}$--$84^{\rm th}$ percentiles. Lower thresholds merge basins more aggressively and therefore produce larger characteristic bubbles, with the separation becoming most apparent during late percolation. The fiducial $10\%$ result remains within the continuous envelope spanned by the neighboring threshold choices.}
    \label{fig:threshold_evolution}
\end{figure}

Figure~\ref{fig:threshold_evolution} shows that the qualitative evolution of the characteristic bubble scale is unchanged across the full threshold range. All choices yield sub-cMpc to cMpc scales at early times followed by rapid growth during percolation. The expected ordering is systematic: smaller $f_{\rm merge}$ values remove more watershed boundaries and shift the volume-weighted distribution toward larger connected regions. The curves diverge most strongly near the end of reionization, when the volume is carried by only a few percolating structures and the treatment of any one interface can appreciably change a volume-weighted statistic.

\begin{figure}
    \centering
    \includegraphics[width=\myfigwidth]{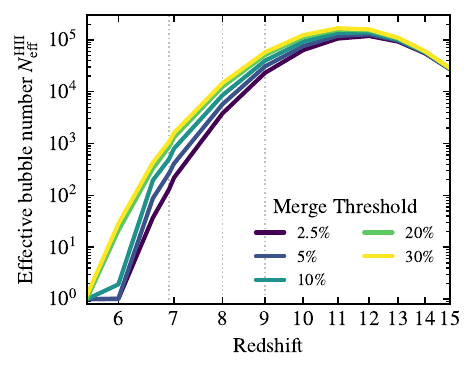}
    \caption[Effective bubble number in the threshold sweep]{Evolution of the effective bubble number for the five shared-surface merge thresholds. Toward lower redshift, the declining effective number indicates fewer effectively contributing bubbles, reducing support for volume-weighted measurements and increasing their sensitivity to the partition of large connected regions. Vertical dotted lines mark $z=9$, $8$, and $6.9$, the epochs selected for the detailed comparisons in Figure~\ref{fig:threshold_distributions} while the effective population remains substantial.}
    \label{fig:threshold_neff}
\end{figure}

We quantify this concentration using the dimensionless effective bubble number defined in Equation~\ref{eq:neff} and shown in Figure~\ref{fig:threshold_neff}. This inverse-participation statistic measures how many comparably important bubbles contribute to a volume-weighted measurement, without implying that they are spatially independent. At $z=9$ and $z=8$, $N_{\rm eff}^{\HII}$ remains in the thousands or above for all five thresholds. At $z=6.9$, it ranges from approximately $130$ to $10^3$ across the threshold sweep, falling to approximately $1$--$30$ by $z=6$. This decline reflects the concentration of ionized volume into a few dominant regions, whose partition increasingly influences the volume-weighted statistics.

We therefore compare the size and shape distributions at $z=9$, $8$, and $6.9$, corresponding to global volume-weighted ionized fractions $x_{\HII}=0.113$, $0.231$, and $0.491$, respectively. These epochs span early bubble growth through the midpoint of reionization and allow us to assess threshold sensitivity before a few large structures dominate the volume-weighted statistics.

\begin{figure*}
    \centering
    \includegraphics[width=\linewidth]{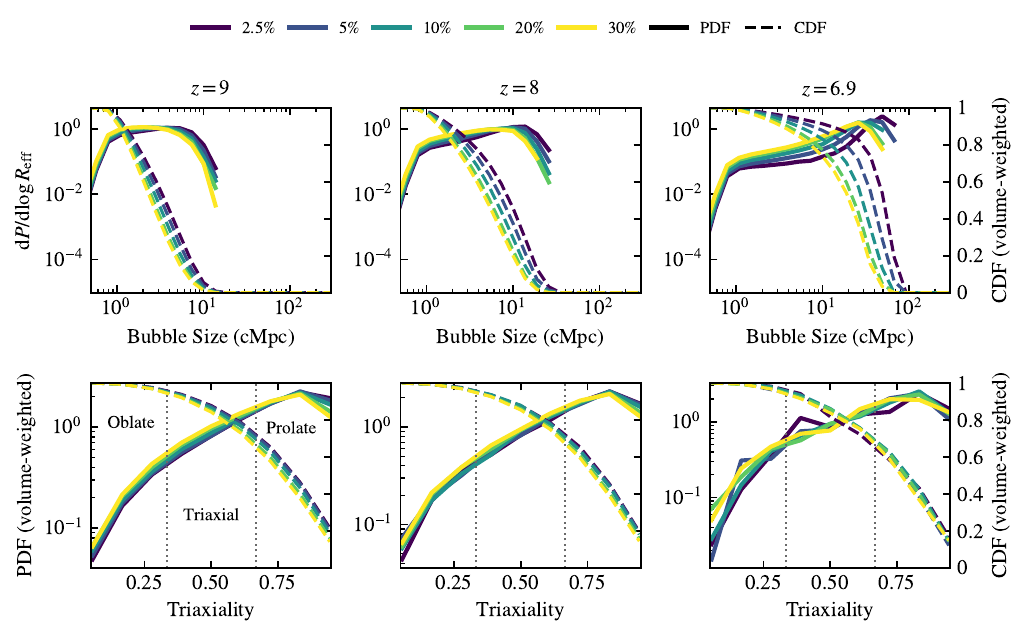}
    \caption[Size and shape distributions in the threshold sweep]{Sensitivity of the volume-weighted bubble-size distributions (upper row) and triaxiality distributions (lower row) to the shared-surface merging threshold at $z=9$, $8$, and $6.9$. Solid curves show probability density functions and dashed curves show cumulative fractions (CDFs), giving the fraction of ionized volume above the indicated radius or triaxiality. Colors denote merger thresholds from $2.5\%$ to $30\%$, with $10\%$ our fiducial choice. The size distributions shift smoothly as more or fewer watershed interfaces are removed. The shape distributions are nearly threshold-independent at $z=9$ and $z=8$ and show moderate variation at $z=6.9$, without a qualitative change in the predominantly prolate morphology.}
    \label{fig:threshold_distributions}
\end{figure*}

Figure~\ref{fig:threshold_distributions} shows that the threshold primarily controls the degree to which neighboring watershed volumes are combined, without introducing qualitatively different populations. The BSDs move progressively toward smaller radii as $f_{\rm merge}$ increases, but retain the same broad form and evolutionary ordering. The triaxiality PDFs and CDFs are especially stable at $z=9$ and $z=8$. At $z=6.9$, some threshold dependence is visible in individual bins, as expected from the lower $N_{\rm eff}^{\HII}$, but the curves vary smoothly and remain concentrated in the prolate regime. We conclude that $f_{\rm merge}=10\%$ is a representative fiducial choice and that the main size and morphology trends reported in this work are robust during the statistically well-sampled portion of reionization. At later times, stronger threshold sensitivity accompanies the decline in $N_{\rm eff}^{\HII}$ as the ionized phase approaches a single percolating network.

\end{appendix}

\end{document}